\documentstyle[12pt,aaspp4]{article}

\newcommand{\simgreat}{\stackrel{\scriptstyle >}{\scriptstyle \sim}}

\lefthead {Geller et al.}
\righthead {The Century Survey}

\begin {document}
\title{The Century Survey: A Deeper Slice of the Universe
\altaffilmark{1,2}}
\author{Margaret J. Geller, Michael J. Kurtz,}
\affil{Harvard-Smithsonian Center for Astrophysics, Cambridge, MA 02138}
\author{Gary Wegner, John R. Thorstensen,}
\affil{Department of Physics and Astronomy, Dartmouth College,
Hanover, NH 03755-3528}
\author{Daniel G. Fabricant,} 
\affil {Harvard-Smithsonian Center for Astrophysics, Cambridge., MA 02138}
\author{Ronald O. Marzke,\altaffilmark{3}}
\affil{Dominion Astrophysical Observatory, Victoria BC V8X 4M6 Canada}
\author{John P. Huchra, Rudolph E.
Schild, and Emilio E. Falco}
\affil{Harvard-Smithsonian Center for Astrophysics, Cambridge, MA 02138}
\altaffiltext{1} {Work reported here based partly on observations obtained at the
Michigan-Dartmouth-MIT Observatory}
\altaffiltext{2} {Work reported here based partly on observations at the
Multiple Mirror Telescope, a joint facility of the Smithsonian Institution
and the University of Arizona}
\altaffiltext{3}{Now at Observatories of the Carnegie Institution of Washington,
Pasadena, CA 91101}
\begin {abstract}

The ``Century Survey'' (CS hereafter) is a complete redshift survey  of
a 1$^\circ$-wide strip. It covers 0.03
steradians to a limiting m$_R$ = 16.13. The survey is 98.4\% complete and
contains 1762 galaxies. Large-scale features in the survey are
qualitatively similar to
those in other surveys: there are large voids surrounded
or nearly surrounded by thin dense regions
which are sections of structures like (and including) the Great Wall.

The survey crosses the classical Corona Borealis supercluster. The galaxy
density enhancement associated with this system extends for $\simgreat 100
h^{-1}$ Mpc (the Hubble constant is H$_0 = 100h$\ km s$^{-1}$Mpc$^{-1}$). 

The Schechter (1976) luminosity function  parameters for the CS are:  
$M^*_{CS} = -20.73 ^{+0.17}_{-0.18}$,
$\alpha_{CS} = -1.17 ^{+0.19}_{-0.19}$, and 
$\phi^*_{CS} = 0.0250\pm0.0061$\ Mpc$^{-3}$mag$^{-1}$.
In concert with the ESO Key Program (\cite{vet97}; \cite{zuc97}) and
the AUTOFIB (\cite{ell96}) surveys, the CS indicates that the absolute
normalization of the  luminosity function exceeds estimates based on
shallower and/or sparser surveys.

\end {abstract}
\keywords { Galaxies: Surveys\ \  Galaxies: Distances and redshifts Galaxies:
Luminosity function\ \  Cosmology: Observations
\ \ Cosmology: Large-Scale structure of the universe}
\section {Introduction}

The ``Century Survey'' (CS) is a complete redshift survey of
102 square degrees to a limiting m$_R$ = 16.13 containing
1762 galaxies. 
It complements the suite of sparse surveys to
comparable or greater depth (\cite{lov92}: 
APM ;\cite{she96} : LCRS ).
The sample size is comparable with that of the much deeper, complete
ESO Key Program of 23.3 square degrees (\cite{vet97}; \cite{zuc97} : ESO-KP). 
The CS covers the central 1$^\circ$-wide declination range of 
the first slice of the CfA survey
(\cite{del86}).
In the northern hemisphere, the CS is the largest
complete survey to its depth.

\begin{figure}
     \figurenum{Plate 1}
     \plotone{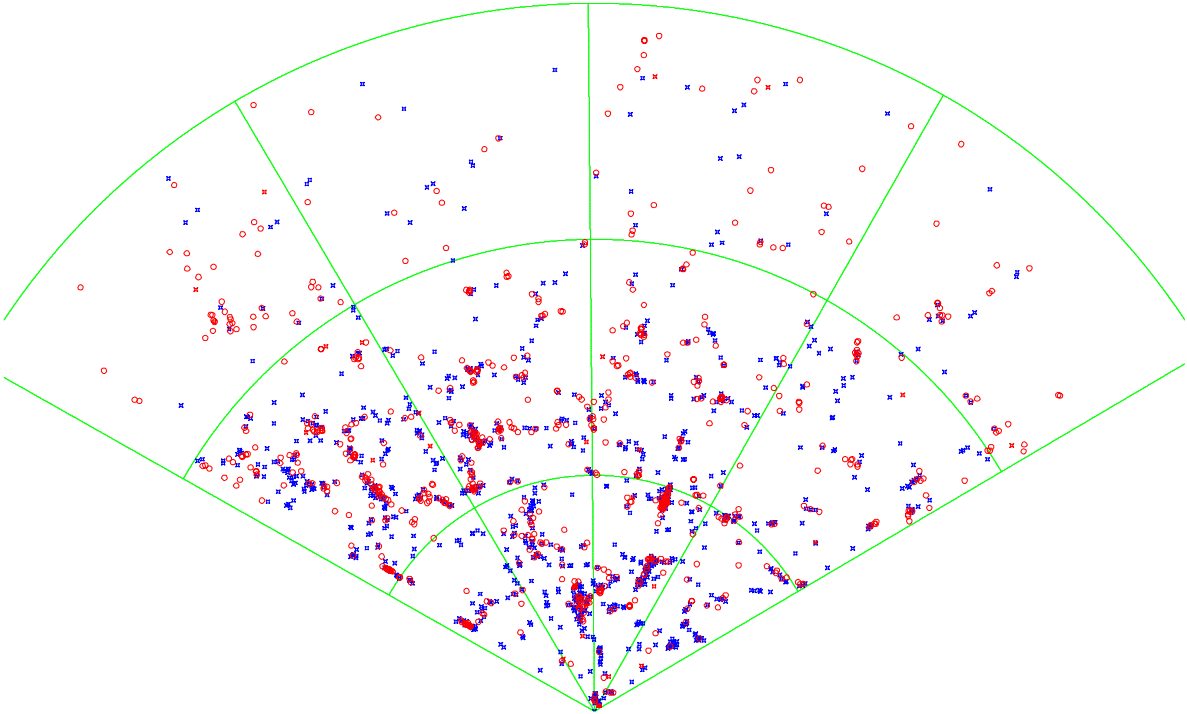}
     \caption{Cone diagram for the Century Survey. The right ascension runs
from 8.5$^{\rm h}$ to 16.5$^{\rm h}$ and the radial green lines are at 2 hour intervals.
The outer boundary of the plot is at 45,000km s$^{-1}$ and the green lines
at constant velocity mark 15,000 km s$^{-1}$ intervals. Blue points
represent spiral galaxies; yellow ones are early types. In the
gray-green regions there
is a mix of types.}

     \end{figure}

The qualitative features of the survey (Plate 1) are similar to the
structures in other large redshift surveys (\cite{gel89}; \cite
{gio89};
\cite{dac94}; \cite{she96}; \cite{vet97}). The
survey slices through a number of large voids and dense, coherent, thin walls
(including a portion of the Great Wall).
An outer region of the Coma cluster ($\alpha = 12^{\rm h}57.5^{\rm m}, +28^\circ
15^\prime$ , $cz = 6850$ km 
s$^{-1}$; \cite {col96}) and a portion of the classical Corona Borealis
supercluster ($\alpha = 15.3 - 15.6^{\rm h}, \delta = 27.5 - 32^\circ$;
see \cite{pos88}; \cite{sma97}) lie
within the CS: these regions contribute prominent ``fingers''
in redshift space. Note that throughout this paper we use B1950 coordinates.

Like the CfA and SSRS2 surveys, the CS is complete  to its magnitude
limit. Although they cover more than a third of the sky, the CfA and
SSRS2 surveys contain structures which extend across most of the
sample volume. One might expect that because of the greater depth of the CS,
individual large features (like the Great Wall) would be less likely to dominate
its large-scale structure statistics. However, the 12,800
km s$^{-1}$ scale claimed by Broadhurst et al. (1990) and the excess power
on a scale of $\sim 100 h^{-1}$ Mpc in the LCRS (\cite{lan96})
suggest caution (see Section 2b). In fact, within the CS, we find further
evidence for inhomogeneity on a $\sim 100 h^{-1}$ Mpc scale.

Here we display the data and discuss the first analyses of the CS.
Section 2 describes the survey. In Section 3 we compute the
$R$-band luminosity function for the entire survey.
We also examine the 
variation  of the
galaxy number density with redshift.
We conclude in Section 4.

\section {The Galaxy Distribution}

\subsection  {The Data}

The CS is a complete photometric and spectroscopic survey of the
region $8^{\rm h}30^{\rm m} \leq \alpha \leq 16^{\rm h} 20^{\rm m}$ and
$29^\circ \leq \delta \leq 30^\circ$. The spectroscopic survey is
98.4\% complete to $m_R = 16.13$ over the entire right ascension range.
We are 99.4\% complete to $m_R = 16.4$ over the right ascension range
$8^{\rm h}32^{\rm m}$ to $10^{\rm h}45^{\rm m}$. In this deeper
sample, there are 518 galaxies; 177 of these have $m_R \leq 16.13$. Here
we discuss the sample complete to $m_R = 16.13$. 
Kurtz et al. (1997) discuss the deeper sample.

We constructed the galaxy catalog from scans of the POSS E plates
according to the procedures outlined
by Kurtz et al. (1985).  For each galaxy in the catalog 
we derive an isophotal magnitude to
a  bright limiting isophote which varies unavoidably from plate to plate.
Two drift
scans from $8^{\rm h}27^{\rm m}$ to $11^{\rm h}55^{\rm m}55^{\rm s}$
and from $11^{\rm h}50^{\rm m}$ to $15^{\rm h}45^{\rm m}$ provide the magnitude calibration
(\cite{ram95}; \cite{ken93}).
The drift scans are both centered at $\delta = 29.5^\circ$.
The drift scan for early $\alpha$'s was done with the 1.2-m
telescope and for late
$\alpha$'s with the 61-cm telescope (now retired) of F. L. Whipple Observatory
(FLWO). We used pointed observations to calibrate the three POSS
plates E924, E1365, and E134 which cover the right ascension ranges
$8^{\rm h}32^{\rm m}32^{\rm s}$
to $8^{\rm h}58^{\rm m}50^{\rm s}$ and
$15^{\rm h}53^{\rm m}$ to $16^{\rm h}19^{\rm m}44^{\rm s}$ (see the next
paragraph).
Calibration details are in Kurtz et al.
(1997).

     \begin{figure}
     \plotone{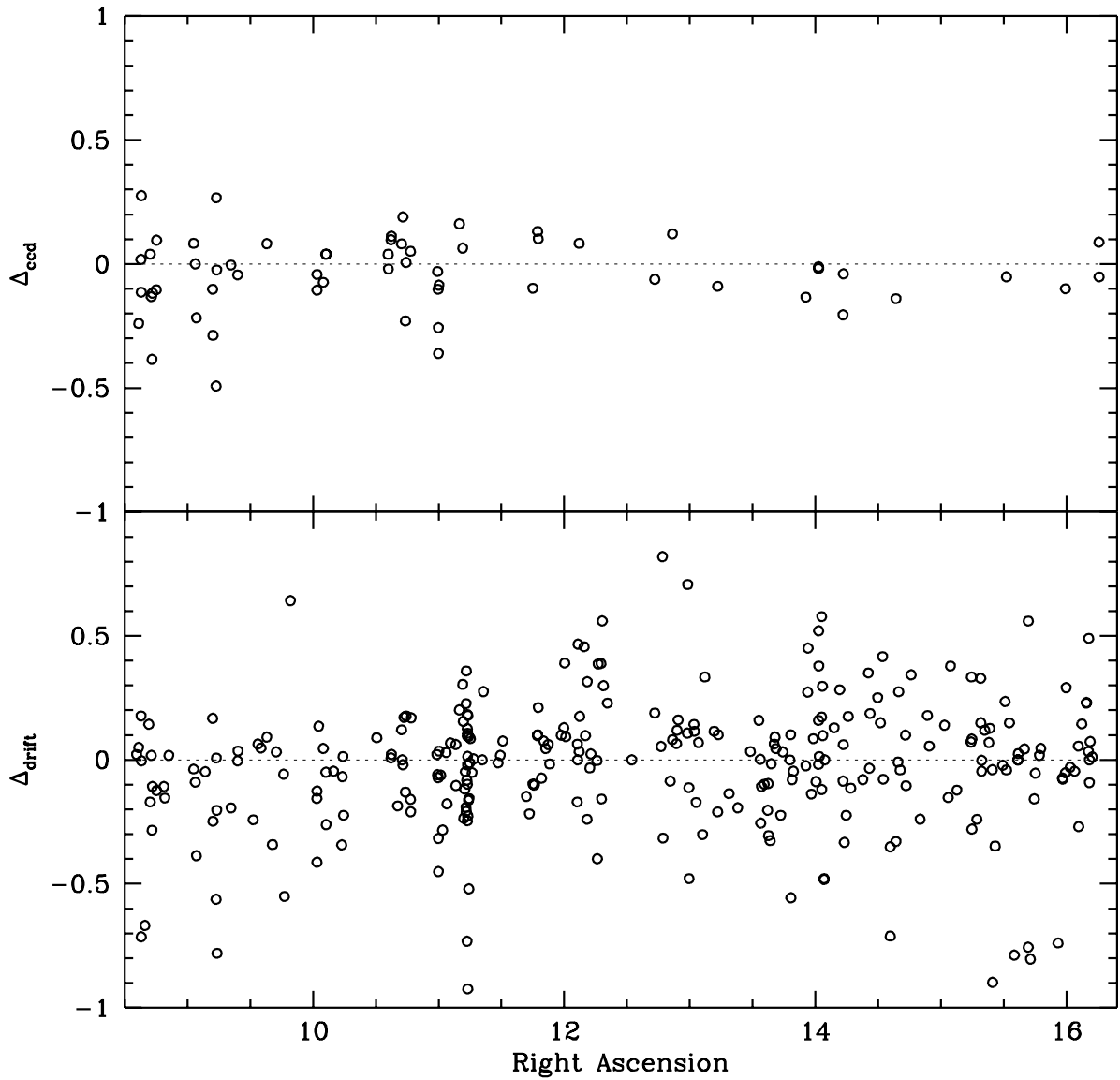}
     \caption{CS photometry. The upper panel
shows the difference between the pointed photometric observations and the
drift-scan calibrated CS as a function of right ascension. The lower panel
shows the difference between the drift scan magnitudes and the drift-scan
calibrated CS as a function of right ascension. Note that
we do not use the drift scan data for calibration in the right ascension
ranges $8^{\rm h}32^{\rm m}32^{\rm s}$ to $8^{\rm h}58^{\rm m}50^{\rm s}$
and $15^{\rm h}53^{\rm m}$ to $16^{\rm h}19^{\rm m}44^{\rm s}$.}

     \end{figure}

As a function of right ascension, Figure 1 (lower panel) shows the difference, 
$\Delta_{drift} = m_{D} - m_{DCCS}$, between the drift
scan magnitudes, $m_{D}$, and the drift-scan calibrated CS,
$ m_{DCCS}$, for the 457 galaxies in common. To calibrate
each plate, we set the median $\Delta_d = 0$. The overall  average is 
$\Delta_{drift} = 0.018 \pm 0.014$ and the rms scatter about the mean is $0.29^{\rm m}$.

To check the drift scan calibration we acquired more than 200 pointed
observations with the MDM 2.4-m and 1.3-m telescopes
under photometric conditions. The photometry
for these
observations is better than 3\% in all cases.
Comparison of these large aperture pointed magnitude
measurements with the drift scans
demonstrates that there are no systematic zero-point variations 
across either drift (see Kurtz et al. 1997 for further details). Figure 1 (upper panel) shows the
difference between 92
pointed photometric observations and the drift-scan calibrated CS as
a function of right ascension.
Except for three plates
noted above (where we calibrate with these
pointed observations only), the pointed photometry provides an independent
check of the CS drift scan calibration.
The rms scatter in $\Delta_{ccd} = m_{p} - m_{DCCS}$ is
0.19$^{\rm m}$ and the median
zero-point offset is $0.015^{\rm m}\pm 0.02^{\rm m}$.
The $m_p$'s are the magnitudes for the pointed
observations.  The scatter here is an underestimate of the error
in the CS photometry because we have not made a detailed accounting of the
photometric errors for objects which are not cleanly separated from their
neighbors. Based on these considerations, we take 0.25$^{\rm m}$ as the 
standard error in
the CS magnitudes.

There are 1762 galaxies with m$_R \leq 16.13$. Only 28 of these do not have
measured redshifts. Most of the galaxies without redshift measurements
are near the limiting magnitude and their inclusion in the survey is largely
a result of final refinement of the magnitude calibration.
We measured  1467 new redshifts;
we measured 631 of these with the Decaspec (\cite{fab90})
mounted on the MDM 2.4-meter. In sparser
regions, where the ten-fiber probe Decaspec would have been
ineffective, we measured 774 redshifts 
either with the blue channel of
the MMT spectrograph (with image intensified Reticon detectors) or
with the red or blue channel
MMT spectrograph (equipped with CCD detectors). The
remaining redshifts are from single-slit 2.4-meter measurements
or from  FAST spectrograph (\cite{fab97}) observations at the 1.5-meter Tillinghast
at FLWO. 
Previously published redshifts in this region are from
Huchra et al. (1990; 164 redshifts), Thorstensen et al. (1989;
92 redshifts), and Willmer et al. (1996; 11 redshifts).

We extracted redshifts from all of the absorption-line spectra
according to the cross-correlation procedure of Tonry and Davis (1979).
For emission-line spectra from MDM and from the MMT spectrograph,
we derive redshifts with a multiple-gaussian fitting routine.  
For the more recent MMT red and blue channel data, we used
the cross-correlation technique for both emission- and absorption-line
spectra. At Dartmouth, we used the REDUCE/INTERACT package
(\cite{mak92}) to extract redshifts; at the CfA
we used EMSAO and XCSAO (\cite{kur97a}).
For all of the data, the typical  error in the velocities 
is 70 km s$^{-1}$. From 67 measurements in common, the zero-point offset
between the MDM and MMT data is $1.2\pm 6.3$ km s$^{-1}$.

One of us (G. Wegner) used a 30 power loupe
to obtain Hubble types from glass copies of the POSS1 O-plates
(the coarse classification bins are
E, S0, Sa, Sb, Sc, Irr and the corresponding
barred types) for
all of the galaxies with m$_R \leq 16.13$. From galaxies
which appear on more than one plate,
we estimate that the classification error is $\pm$1 type. 
The median type in this $R$-selected survey is Sa.
We describe and analyze the classifications in much greater detail
in Kurtz et al. (1977).

We make two corrections to the
computed absolute magnitudes: 

$$
M(m,z,T) = m-5\log D_L(z)-25-K_T(z)-A_R
$$

\noindent 
where m is the apparent $R$ magnitude from the photometric catalog, 
$D_L$ is the luminosity distance in megaparsecs at redshift $z$, 
$K_T$ is the $R$-band $K$-correction
for Hubble type $T$, and $A_R$ is the extinction at $R$.
We correct for Galactic extinction using the
reddening maps of Burstein and Heiles (1982)
and a ratio of total to selective absorption of $R_R=2.3$. 
The majority of the survey galaxies lie at high Galactic latitude;
the largest absorption correction is 0.09 magnitudes.
We determine K-corrections for each galaxy type using the spectral
synthesis models of Rocca-Volmerange and Guiderdoni (1988). At the
chosen redshift limit of the survey, $cz=45,000 \rm \,km \, s^{-1}$, these K-corrections
range from $-0.02$ magnitudes for irregulars to 0.2 magnitudes for
ellipticals.
We compute distances directly from the galaxy redshifts in the Local Group
frame: 

\begin{eqnarray*}
D_L(z) & = & {c \over H_0}(1+z) Z_q(z) \\
Z_q(z) & = & {1 \over q_0^2(1+z)}\lbrace q_0z+(q_0-1)\lbrack (1+2
q_0z)^{1/2}-1 \rbrack \rbrace \\
\end{eqnarray*}

\noindent where
$cz = cz_\odot + 300\ {\rm sin}l\ {\rm cos}b$.  To facilitate
comparisons with earlier work, we use
$H_0 = 100 h$ km s$^{-1}$Mpc$^{-1}$ with $h = 1$. 
We assume a mass density
$\Omega_0=0.3$ (consistent
with several recent measurements) and zero cosmological constant:
thus the deceleration parameter is $q_0 = 0.15$.

\subsection {The Redshift Map}

Plate 1 shows the distribution of the CS galaxies with $cz \leq 45,000$
km s$^{-1}$ color-coded by morphological type. Blue signifies types Sa and later;
yellow corresponds to the early types E and S0. Regions occupied by both
early and late types are gray-green. 

Three qualitative results are evident in the map. First, both
early- and late-type galaxies trace the large-scale features in the
survey. This result is also evident in the two slices of the CfA2
survey for which types are available (\cite{huc90}; \cite{huc95}).
Second, the
few galaxies within the lowest
density regions (voids) tend to be of late-type, but there are {\it some}
\ E's and S0's.  In a shallower survey of the
first CfA slice, Thorstensen et al. (1995) also noted
the presence of
early-type galaxies in low density regions. 

Finally, the cores of rich
clusters are rich in early-type galaxies.  Portions
of the central fingers appear dark green
in Plate 1 because the cores are surrounded by late-type galaxies
(a few are within the core)  which
generally have a more extended distribution also visible in Plate 1.
The superposition of early and late types at the same position produces
the gray-green dots. These
effects are already well known from detailed studies of individual
clusters (\cite{col96}; \cite{moh96}; \cite{car97}).

The CS contains  portions of 7 Abell clusters (A690, A1185,
A1213, A1656, A2079, A2162, and A2175). The prominent ``finger'' at
$\sim 13^{\rm h}$ is part of the Coma cluster; its center is south of the survey
boundary at $28^\circ 15^\prime$.

The cluster A2079
($15^{\rm h}26^{\rm m}$, $29^\circ 15^\prime$, $cz = 19,600$\ km 
s$^{-1}$) is one of the six clusters in the 
classical Corona Borealis supercluster (Cor Bor). The other five clusters
(A2061, A2065, A2067, A2089, and A2092) are outside the declination range of
the CS. These systems span the right ascension range $15.3^{\rm h} - 15.6^{\rm h}$
and the approximate velocity range 18,000 - 25,000 km s$^{-1}$
(\cite{pos88}; \cite {sma97}). In this velocity range, 
the CS is surprisingly
dense all the way from 14.5$^{\rm h}$ to 16.5$^{\rm h}$.
At the median redshift of  Cor Bor 
($cz \sim 22,000$ km s$^{-1}$, which also happens to be the depth, D$^*$, to which an
L$^*$ galaxy is brighter than the magnitude limit), the
dense region has an extent of
$\sim 100 h^{-1}$ Mpc , comparable with the scale where excess
power is detected in the Las Campanas Redshift Survey (LCRS; \cite{lan96}).

Counts in the cells marked in Plate 1 give a measure of the
extent of Cor Bor and of its contrast
with other portions of the CS. In the
easternmost cell with velocity range 15,000 to 30,000 km s$^{-1}$,
there are 348 galaxies (only 92 of these are in the canonical $15.3 - 15.6^{\rm h}$
Cor Bor range). In  this velocity range
the counts in the other three cells from east to west are,
respectively: 214, 184, and 141. In the cell which contains Cor Bor, the
galaxy number density is $\sim 1.7$ times the overall median.
The east-west gradient does not persist
throughout the velocity range of the survey
and thus it is not a result of an undiscovered systematic error in the
photometry.
In fact, we compute
(Section 3) the
luminosity function parameters for the eastern and western halves of the
survey; all three parameters (the amplitude, characteristic luminosity,
and faint-end slope) are indistinguishable at the 1$\sigma$ level.

\section {The Luminosity Function}

There are  several recent surveys   containing $\simgreat$ 1000 galaxies
which yield determinations of the local luminosity function. They include
Loveday et al. (1992: APM), Marzke et al.
(1994a,b: CfA2), da Costa et al. (1994: SSRS2), 
Shectman et al. (1996: LCRS),
Ellis et al. (1996: AUTOFIB), and  Zucca et al. (1997: ESO-KP). The  APM, CfA2,
SSRS2,  AUTOFIB, and ESOKP are all blue-selected surveys. Like the CS, the LCRS
is red-selected (at Gunn $r$ rather than at the Kron-Cousins $R$ we use;
$r - R = 0.35$; Jorgensen 1994). There should be some variation in the
{\it shape} of the luminosity function with
color because of the expected associated change in morphological mix. The
CS luminosity function parameters are thus
most directly comparable with the LCRS. For comparison with
$b_J$-selected surveys, the median is $(b_J - R_{KC}) = 1.3$ for the CS
mix of morphological types (\cite{but95a}; \cite{but95b}).

We use the STY maximum-likelihood
(\cite{sand79}; \cite{yah91})
and stepwise maximum
likelihood (\cite{efs88} : SWML) techniques to
determine the parameters M$^*$ and $\alpha$ in the Schechter (1976) function:
\begin {equation}
 \phi{(M)} = 0.4 ln10\ \phi^*[10^{0.4(M^*-M)}]^{(1+\alpha)}{\rm exp}
[10^{0.4(M^*-M)}]. 
\end {equation}

\noindent Both methods are inhomogeneity-independent. We assume
that the luminosity function is independent of position. Thus
we can determine its form and amplitude, $\phi^*$, separately. The STY
method is parametric; the SWML method provides a measure of the goodness
of fit for the STY parameters, $M^*$ and $\alpha$.

The {\it observed} luminosity
function $\phi{(M)}$ is a
convolution of the magnitude error
distribution with the {\it true} luminosity
function of equation (1).
We assume that the magnitude error distribution
is  Gaussian with a dispersion of $\sigma_M = 0.25$.

We construct the luminosity function for
$1,000\ {\rm km\ s^{-1}} \leq cz \leq 45,000\ {\rm km\ s^{-1}}.$
The lower limit eliminates regions where peculiar velocities may be a
large fraction of the cosmological recession velocity; the upper
limit is where the survey becomes too sparse to be useful.

     \begin{figure}
     \plotone{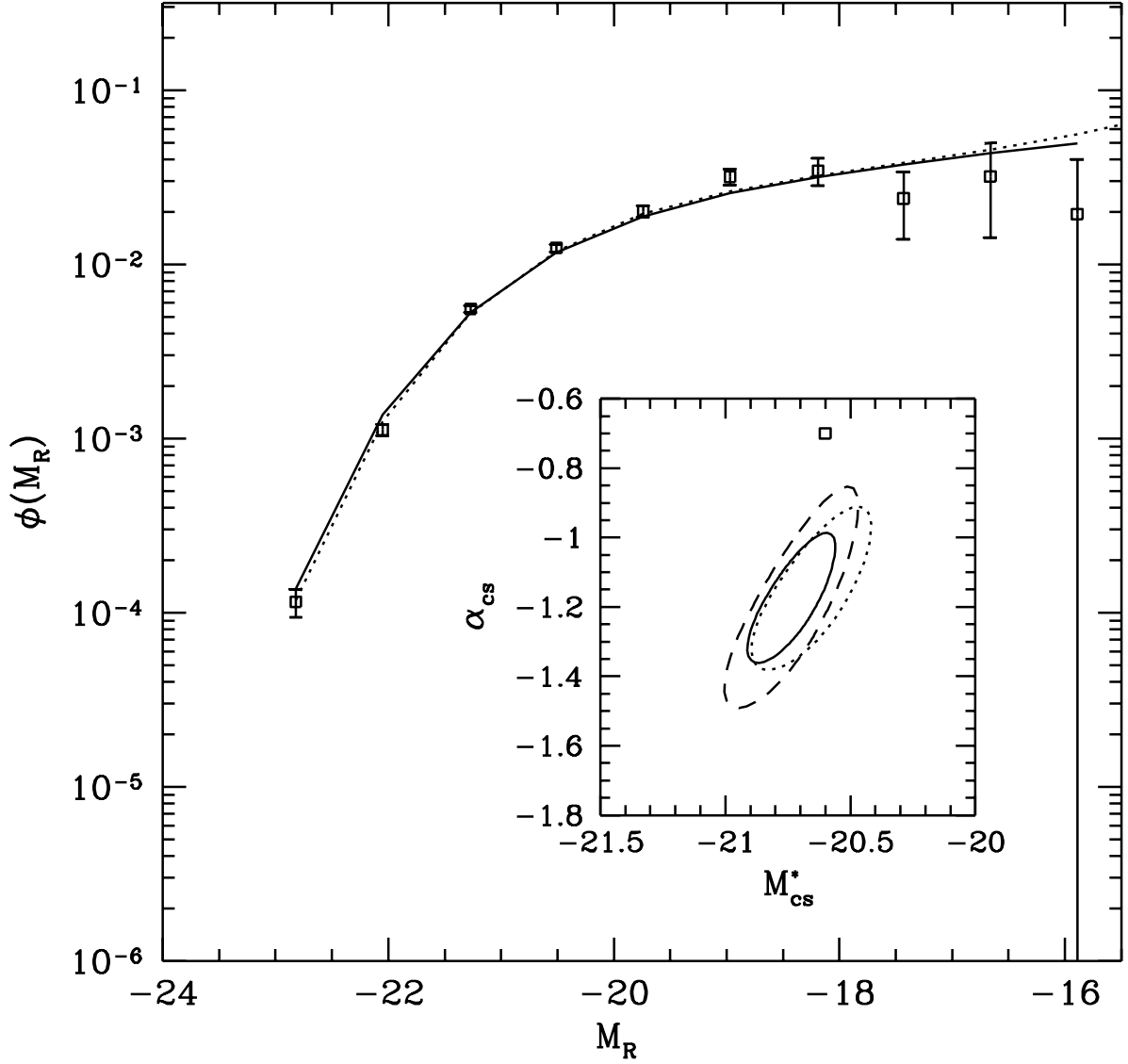}
     \caption{The CS luminosity function. The
points with error bars are the SWML estimates. The solid curve is the STY
fit with the parameters listed in Table 1. The dotted
curve is the $R$-band luminosity function predicted from the $B$-band SSRS2.
The STY fit is based
on galaxies with 1,000 $ < cz \leq 45,000$ km s$^{-1}$. The inset shows
the $1\sigma$ error
ellipses for the luminosity function parameters for the entire survey
(solid), the East half (dashed), and the West half (dotted). The open box shows
the best fit for the LCRS.}
     \end{figure}

Figure 2
shows the binned STY (curve) and SWML (symbols
with error bars) estimates of the {\it observed} 
(convolved) luminosity
function for the entire CS sample. 
Inversion of the information matrix yields the error bars
on the SWML estimates 
(\cite{efs88}).
Marzke et al. (1994a) describe this approach in detail.

The inset in Figure 2 shows the 1$\sigma$ error
ellipse for the Schechter (1976) function
parameters, $M^*_{CS} = -20.73 ^{+0.17}_{-0.18}$ and
$\alpha_{CS} = -1.17 ^{+0.19}_{-0.19}$ derived
from the STY technique. With $r -R = 0.35$, $M^*_{CS}$ is coincident
with $M^*_{LCRS}$ (Table 1; the open box within the inset
shows the best-fit LCRS parameters). However, the CS faint-end slope is
significantly ($\sim 2.5 \sigma$) steeper than $\alpha_{LCRS} = - 0.70\pm
0.05$.  The source of this discrepancy is unclear. One possibility
is that the selection criteria used in the LCRS bias the
survey against the faintest galaxies.  Because of the 
correlation between absolute magnitude and surface brightness,
the central surface-brightness cut in the LCRS is the criterion
most likely to eliminate galaxies from the faint end of the LF.
The  effect may, in fact, be small, but it remains to be evaluated.
Although our survey (as any magnitude-limited survey) 
is biased to high surface brightness galaxies as well (Kurtz et al. 1997), 
we avoid explicit cuts on anything but apparent $R$ magnitude.
This difference in survey strategy between the CS and the
LCRS may contribute to 
the discrepancies at the faint end of the LF.

\begin{deluxetable}{cccccr}
\tablecolumns{6}
\tablewidth{0pt}
\tablecaption{Luminosity Function Parameters}
\tablehead{\colhead{Sample} & \colhead{$N_{gal}$} & \colhead{$M^*(R_{KC})$} &
\colhead{$\alpha$} & \colhead{$\phi^* (\times {10^{-3} \,\rm Mpc^{-3})}$} &
\colhead{notes} }
\startdata

 Century-All & 1695\tablenotemark{a} & $-20.73^{+0.17}_{-0.18}$ & $-1.17^{+0.19}_{-0.19}$
&25.0$\pm$6.1 & \nodata \nl
 Century-East & 846\tablenotemark{a} & $-20.78^{+0.25}_{-0.29}$ & $-1.20^{+0.32}_{-0.31}$
&32.1$\pm$11.1 & \nodata \nl
 Century-West & 849\tablenotemark{a} & $-20.68^{+0.22}_{-0.26}$ & $-1.15^{+0.23}_{-0.24}$
&21.4$\pm$12.2 & \nodata \nl
 LCRS & 18678 & $ -20.64^{+0.02}_{-0.02}$ & $-0.70^{+0.05}_{-0.05}$ & $19.0\pm 1.0$ & $(r-R_{KC}) = 0.35$ \nl
 ESO-KP & 3342 & $-20.91^{+0.06}_{-0.08}$ & $-1.22^{+0.06}_{-0.07}$ & $20.0\pm4.0$ & $(b_j - R_{KC}) = 1.30$\nl
 AUTOFIB & 588 & $-20.60^{+0.15}_{-0.12}$ & $-1.16^{+0.05}_{-0.05}$ & $24.5^{+3.7}_{-3.1}$ & $(b_j - R_{KC}) = 1.30$ \nl
\tablenotetext{a}{$\alpha$ and $M^*$ are based on this sample size; 
$\phi^*$ is based on the subset in the $18,000-28,000 km/s$ range; 
653(All), 246(West), 407(East)}
\enddata
\end{deluxetable}

Although the LCRS slope differs significantly from the one
we measure, comparable surveys in other passbands 
yield results which are quite consistent with ours.  In particular, 
our $\alpha_{CS}$ is indistinguishable from the faint end slopes 
measured in the ESOKP (Zucca et al. 1997) and in the AUTOFIB surveys
(Ellis et al. 1996, for the regime $0.02 < z < 0.15$), 
both of which are selected in $b_J$.

The agreement between our $R$-selected survey and recent
blue-selected surveys would be surprising 
if the early-type luminosity function were radically different from
the late-type luminosity function.
In the very local $B$-selected samples,
the early-type luminosity function is essentially indistinguishable from the 
spiral luminosity function (SSRS2:Marzke et al. 1997, CfA2: Marzke et al. 1994).
At redder wavelengths, the luminosity function of red, early-type galaxies must then
be brighter than the luminosity function of blue, late types. Because late types 
outnumber early types by a factor of nearly two, the 
overall luminosity function develops a feature around the knee in bandpasses
where the color
differential between early and late types 
exceeds a magnitude or so.  In the $R$
band, the color difference between ellipticals and Sbc's is only
half a magnitude.  The net effect is that in the $R$-band, 
the early-type galaxies determine M$^*$; late types still govern the
faint-end. 

The dotted curve in
Figure 2 shows the $R$-band luminosity
function predicted from the $B$-selected
SSRS2 luminosity function. 
Marzke et al. (1997) divide the
SSRS2 luminosity function into three type bins: E/S0, Spiral and Irregular.
The SSRS2 luminosity function parameters are (Marzke et al. 1997) : E/S0 ,
M$^* =-19.42^{+0.10}_{-0.11}$, $\alpha =
-1.03^{+0.09}_{-0.09}$, $\phi^* = 4.3\pm 0.8\times 10^{-3}$ Mpc$^{-3}$;
Spiral,
$M^* = -19.46^{+0.07}_{-0.08}$, $\alpha = -1.11^{+0.07}_{-0.07}$,
$\phi^* = 8.1\pm 1.3\times 10^{-3}$ Mpc$^{-3}$;
Irregular, $M^* -19.72^{+0.41}_{-0.52}$, $\alpha = -1.86^{+0.25}_{-0.26}$,
$\phi^* = 0.2\pm 0.1\times 10^{-3}$ Mpc$^{-3}$.
For a basic model, we translate these $B$-selected luminosity functions to the $R$ band using 
the mean $B-R$ colors from Frei and Gunn (1995). We do not account
for the large intrinsic scatter in the color for each morphology. For ellipticals,
Sbc's (roughly the mean spiral type in SSRS2) and irregulars,
the colors are 1.39, 0.95 and 0.57, respectively. 
We adjust the normalization by the ratio of 
${\phi_*}_{CS}/{\phi_*}_{SSRS2}$  to compare the shapes directly.
The predicted $R$-band luminosity function agrees remarkably well with
the CS luminosity function.

Gardner et al. (1997) determined
the galaxy luminosity function for a K-band survey. Here too, the faint-end
slope, $\alpha = - 0.91 \pm 0.2$, is nearly flat. 
Because late spirals and ellipticals
differ in $B-K$ by more than two magnitudes, the early and
late-type luminosity functions are well separated at $K$. The composite luminosity function may reflect this
separation.

We also compute luminosity functions for the east and west
halves of the photometric CS (the
division is at $\alpha = 12^{\rm h}56^{\rm m}18^{\rm s}$).
To within the  errors, all
of the luminosity function parameters 
for the subsamples and
for the survey as a whole are similar.  The inset of Figure 2 shows the error
ellipses for the East (dashed) and West (dotted) samples. The nearly complete
overlap demonstrates the consistency of the photometry across the
survey. Table 1 lists the number of galaxies used to compute the luminosity
function for each sample.

To compute the luminosity function
amplitude, $\phi^*_{CS}$, we use the minimum variance estimator derived
by Davis and Huchra (1982).  
Figure 3 shows the behavior of $\phi^*_{CS}$ as a function of velocity.
The Cor Bor region (Section 2b) causes the peak at
$\sim$ 24,000 km s$^{-1}$.  The peak centered at 9,000 km s$^{-1}$ is the
Great Wall, bounded
by a few large voids  in  its foreground and background.
Examination of
Plate 1 shows that the fluctuations at $cz \leq 16,000$ km s$^{-1}$
result from a small number of large-scale features.

     \begin{figure}
     \plotone{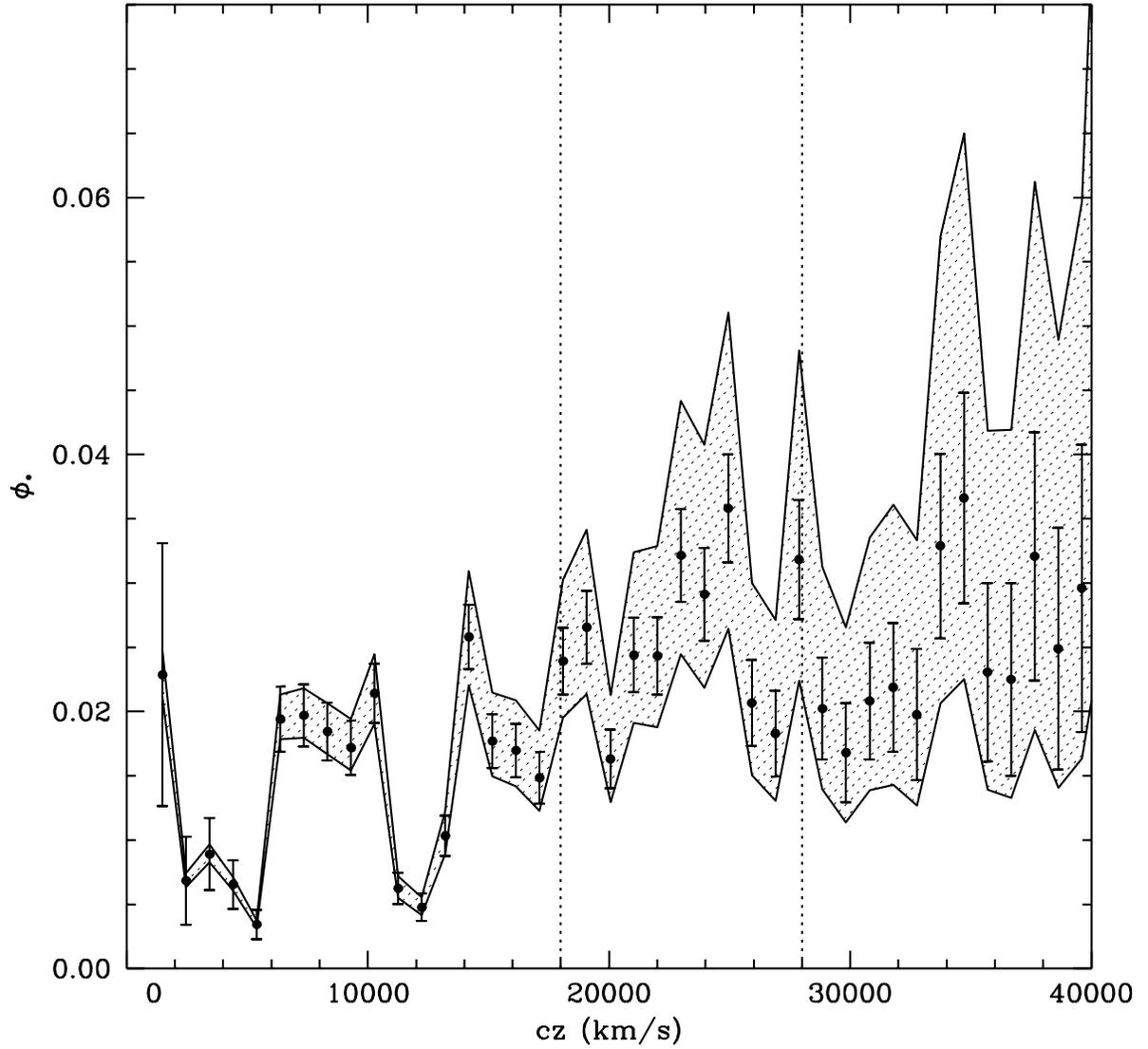}
     \caption{Absolute normalization for the Century Survey luminosity
function. The points with error bars are the best estimates. The
boundaries of the dotted region are the results for the extrema of the error
ellipse in Figure 2. The vertical dotted lines indicate the ranges for
computation of the median $\phi^*$.} 
     \end{figure}

In Figure 3, the points with
error bars are the estimates based on the best-fit luminosity function
parameters for the  entire sample in Table 1. The
error bars are the formal error in the estimator. The boundaries
of the shaded region  show the behavior of $\phi^*_{CS}$ 
at the extrema of the error ellipse in Figure 2.

The luminosity function fit is
most sensitive to the galaxies with
magnitudes near $M^*_{CS}$. In Table 1 we therefore
quote the average $\phi^*_{CS}$ for the velocity range [18,000:28,000] centered
on the velocity (22,000 km s$^{-1}$) where an $M^*_{CS}$ galaxy is at the
magnitude limit of the survey. The vertical
dotted lines in Figure 3 mark this range
which includes 653 galaxies. The result, 
$\phi^*_{CS} = 0.0250\pm 0.0061$\ Mpc$^{-3}$mag$^{-1}$, agrees
very well with the ESOKP and AUTOFIB surveys and is within $1\sigma$ of
the LCRS value. Our error is large
because of the fluctuation induced by Cor Bor
(see Section 2.2).   
The difference in $\phi^*$ between the east and west halves of the sample,
though not formally significant, is in the sense expected from the presence of Cor
Bor in the east half.
 
To examine the dependence of $\phi^*_{CS}$ (or, equivalently, galaxy density) on
velocity, we divide the sample into three velocity bins: [1,000: 18,000],
[18,000: 28,000], and [28,000:45,000] marked in Figure 3. 
The value of $\phi^*_{CS}$ is quite sensitive to the form parameters,
$M^*$ and $\alpha$. 
To examine this sensitivity for each velocity range, we compute the
median value of $\phi^*_{CS}$ and
of the boundaries of the 1$\sigma$ error region for the form parameters
(Figure 2).  The values of $M^*$ and
$\alpha$ are fixed throughout the entire
velocity range. For the
successive velocity intervals,
the ranges of $\phi^*$ are: (0.015, {\bf 0.017}, 0.021); 
(0.021, {\bf 0.027}, 0.034) and (0.014, {\bf 0.025}, 0.049). The boldface
values are the median $\phi^*_{CS}$. The other two numbers are the
median value for the lower and upper boundaries of the shaded region.

These  estimates of $\phi^*_{CS}$ suggest,
albeit with modest significance, that the mean galaxy density is
low nearby and rises by 50\%  for $cz \simgreat 18,000$ km s$^{-1}$. The
amplitude for $cz \simgreat 18,000$ km s$^{-1}$ agrees very well with the
results of the AUTOFIB survey. The behavior of $\phi^*_{CS}$
is similar to that of the ESO-KP (Table 1). On the other hand,
the 1$\sigma$ limits provide a cautionary note.
They show that the CS admits either 
a constant density or a steadily increasing one and underscore the
sensitivity of the behavior of $\phi^*$ to the form parameters of the
luminosity function. Any survey errors in M$^*$ comparable with the CS
errors would presumably show similar behavior. 

The CS samples a region of
the universe completely disjoint from those sampled by the other surveys to
comparable depth.
Like  other recent redshift surveys to comparable depth, the CS indicates
that the sparse APM survey and 
shallower surveys like the SSRS2 underestimate the absolute normalization
of the galaxy luminosity function.
This local
underestimate probably results from large-scale structure, but
the details remain unclear. The CS (and other surveys) do show that
there {\it are} galaxy density enhancements (like Cor Bor) which exceed 
the mean  galaxy density by  $\sim 70\%$ on a scale of 
$\sim 100 h^{-1}$ Mpc. A local underdensity on a comparable scale would
explain the results.

\section {Conclusion}

Comparison of the CS luminosity function with other surveys presents an
interesting, if somewhat puzzling, picture. The luminosity function
parameters, M$^*$ and $\phi^*$,
are in  agreement with the LCRS and with dense $b_J$-selected surveys to
comparable depth. 
However, the faint-end slope for the LCRS is shallower  than
the CS, and the $b_J$-selected surveys.
It is encouraging that we can predict the CS $R$-band luminosity function from
the much shallower SSRS2 $B$-band survey.

The CS  is complete
to the magnitude limit and requires no sampling corrections. For
the other surveys listed in Table 1, the luminosity function
computation requires some correction for incompleteness. These corrections
are smallest for the ESO-KP and AUTOFIB.
In concert with the ESO-KP and AUTOFIB surveys, the CS indicates that the absolute
normalization of the  luminosity function exceeds estimates based on
shallower and/or sparser surveys.

The LCRS indicates that there are inhomogeneities in the galaxy
distribution on a scale of 100$h^{-1}$ Mpc. The Century Survey covers
too small a volume to provide a comparable measure, but it does cross
the Corona Borealis supercluster. This system is embedded in
a dense region much more spatially extensive than the classical supercluster.
The galaxy number density exceeds the mean by $\sim 70\%$ on a scale of $\sim
100h^{-1}$ Mpc. Underdense regions on this scale might account for
the low estimates of the luminosity function normalization at
low redshift.

The CS data include spectroscopic types and galaxy morphologies. 
Kurtz et al. (1997) will discuss  the dependence
of the luminosity
function on these characteristics. This paper will also include an analysis
of the portion of the CS sample with m$_R \leq 16.4$. Subsequent papers
on the CS will include 1) an analysis of the low-order
moments of the pairwise velocity
distribution and
2) a data paper which includes all of the spectroscopic and photometric
data.  

\acknowledgements
We thank Steve Kent and Massimo Ramella for providing their drift scan data 
to calibrate our survey. Mario Nonino generously helped us to understand the
drift-scan data. Carol Heller, John McAfee, and Janet Miller
provided expert assistance in making
the MMT observations. We thank Shoko Sakai for making some of the
MDM 2.4-meter observations and Bob Barr for his excellent support
at the 2.4-meter.
We thank Susan Tokarz for reducing the MMT and 1.5-meter data
and we thank Jim Peters and Perry Berlind for making the
1.5-meter observations.  We thank Peter Challis and Robert Kirshner
for measuring a small number of redshifts
at the MMT to help make the survey complete. We thank the referee, Richard
Ellis, for leading us to consider the relationship between
R- and B-band luminosity functions in more detail.

At the CfA, this research
was supported in part by NASA Grant NAGW-201 and by the Smithsonian
Institution. We thank Irwin Shapiro for providing substantial funding
for the construction of the Decaspec at the CfA. Edward Hertz did the
skillful, dedicated Decaspec engineering. At
Dartmouth, JRT was supported in part by NSF grant AST
86-20081 and by a Research Corporation Cottrell Grant; GW was supported in
part by NSF grants AST86-20081, AST90-23450, and AST93-47714. Dean Bruce
Pipes provided partial financial support for
the construction of the Decaspec.

\clearpage


\clearpage   

\begin{thebibliography}{}



\bibitem [Broadhurst et al.,\ 1990]{bro90} Broadhurst, T.J.,
Ellis, R.S., Koo, D., Szalay, A.S. 1990, 
\nat ,  343, 726.
\bibitem [Burstein and Heiles 1982]{bur82} Burstein, D.
and Heiles, C. 1982, \aj , 87, 1165. 
\bibitem [Buta et al.,\ 1995]{but95a} Buta, R.,
Corwin, H.G., de Vaucouleurs, G., de Vaucouleurs, A.,
Longo, G. 1995, \aj , 109, 517.
\bibitem [Buta and Williams 1995]{but95b}Buta, R. and Williams, K.L. 1995, \aj , 109, 543.
\bibitem [Carlberg et al., \ 1997]{car97}Carlberg, R.G., Yee, H.K.C.,
Ellingson, E., Morris, S.L., Abraham, R., Gravel, P., Pritchet, C.J., 
Smecker-Hane, T., Hartwick, F.D.A., Hesser, J.E., Hutchings, J.B. and Oke,
J.B. 1997, \apjl , 476, L7
\bibitem [Coleman et al., \ 1980]{col80}Coleman, G.D., Wu, C-C. and Weedman,
D.W. 1980, \apj , 43, 393.
\bibitem [Colless and Dunn 1996]{col96} Colless, M. and
Dunn, A.M. 1996, \apj , 458, 435.
\bibitem [da Costa et al.,\  1994]{dac94} da Costa, L.N.,
Geller, M.J., Pellegrini, P.S., Latham, D.W., Fairall, A.P.,
Marzke, R.O., Willmer, C.N.A., Huchra, J.P., Calderon, A.P., Ramella, M., and
Kurtz, M.J. 1994, \apjl , 424, L1.


\bibitem [Davis and Huchra 1982] {dav82a} Davis, M.
and Huchra, J.P. 1982, \apj , 254, 437.


\bibitem [de Lapparent et al., \ 1986]{del86} de Lapparent, V. , Geller, M.J., and Huchra, J.P. 1986, \apjl ,
302, L1.
\bibitem [Ellis et al., \ 1996] {ell96}Ellis, R.S.,
Colless, M., Broadhurst, T., Heyl, T. and
Glazebrook, K. 1996, \mnras, 280,235.
\bibitem [Efstathiou et al.,\ 1988] {efs88} Efstathiou, G., Ellis, R,S.,
and Peterson, B.A. 1988, \mnras , 
280, 25.



\bibitem [Fabricant and Hertz 1990]{fab90} Fabricant, D.G.
and Hertz, E. 1990, \procspie , 1235,
747. 
\bibitem [Fabricant et al.,\  1997]{fab97} Fabricant, D.G., Cheimets, P.,
Caldwell, N. and Geary, J. 1997, 
\pasp , in press.
\bibitem [Frei and Gunn 1994]{fre94} Frei, Z.
and Gunn, J.E. 1994,\aj, 108, 1476.

\bibitem [Geller and Huchra 1989]{gel89} Geller, M.J. and
Huchra, J.P. 1989, {\it Science}, 246, 897.

\bibitem [Giovanelli and Haynes 1989]{gio89} Giovanelli, R.
and Haynes, M.P. 1989, \aj , 97, 633  




\bibitem [Huchra et al.,\ 1990]{huc90} Huchra, J.P.,
Geller, M.J., de Lapparent, V. and Corwin, H. 
1990, \apjs ,   72, 433.
\bibitem [Huchra et al.,\ 1995]{huc95} Huchra, J.P. Geller, M.J.
and  Corwin, H.J. 1995,
\apjs , 99,
391. 
\bibitem [Jorgensen 1994]{jor94} Jorgensen, I. 1994, \pasp ,  106, 967.
\bibitem [Kent et al.,\ 1993]{ken93} Kent, S.M.,
Ramella, M. and Nonino, M. 1993, \aj ,  105,
393.
\bibitem [Kurtz et al.,\ 1985]{kur95} Kurtz, M.J.,
Huchra, J.P., Beers, T.C., Geller, M.J., Gioia, I.M.,
Maccacaro, T., Schild, R.E., Stauffer, J.R. 1985, \aj ,  90, 1665.
\bibitem [Kurtz and Mink 1997]{kur97a} Kurtz, M.J. and Mink, D. 1997, in preparation.
\bibitem [Kurtz et al.,\ 1997]{kur97b} Kurtz et al. 1997, in preparation.
\bibitem [Landy et al.,\ 1996]{lan96} Landy, S.D.,
Shectman, S.A., Kirshner, R.P., Oemler, A., Tucker, D.
1996, \apjl , 456, L1.
\bibitem [Lin et al.,\ 1996]{lin96} Lin, H.,
Kirshner, R.P., Shectman, S.A., Landy, S.D., Oemler, A.,
Tucker, D.L., Schechter, P.L. 1996, \apj , 464, 60.
\bibitem [Loveday et al.,\ 1992]{lov92} Loveday, J.,
Peterson, B.A., Efstathiou, G., Maddox, S.J.
1992, \apj , 390, 338.
\bibitem [Maker et al.,\ 1982]{mak92} Maker, S.,
Kurtz, M.J. and La Sala, J. 1982, {\it The
REDUCE/INTERACT Data Reduction System} (Hanover: Dartmouth College
Department of Physics and Astronomy).
\bibitem [Marzke et al. \ 1994a]{mar94a} Marzke, R.O.,
Huchra, J.P., Geller, M.J. 1994, \apj,  428,
43.
\bibitem [Marzke et al., \ 1994b]{mar94b}Marzke, R.O.,
Geller, M.J., Huchra, J.P. and Corwin, H.G. 1994
{\it AJ}, {\bf 108}, 437.

\bibitem [Marzke et al., 1997]{mar97} Marzke, R.O., da Costa, L.N.,
Pellegrini, P.S. \& Willmer, C.N.A. 1997, \apj, submitted.

\bibitem [Mohr et al., \ 1996]{moh96}Mohr, J.J., Geller, M.J., Fabricant,
D.G., Wegner, G., Thorstensen, J.R. and Richstone, D.O. 1996, \apj ,
470, 724.
\bibitem [Postman et al.,\  1988]{pos88} Postman, M.,
Geller, M.J., and Huchra, J.P. 1988, \aj , 
95, 571.

\bibitem [Ramella et al.,\ 1995]{ram95} Ramella, M.,
Nonino, M. and Geller, M.J. 1995, {\it Mem.S.A.I.},
66, 113.

\bibitem [Rocca-Volmerange \& Guiderdoni, 1988]{rvg88} Rocca-Volmerange, B.
\& Guiderdoni, B. 1988, A\&AS, 75, 93.

\bibitem [Sandage et al.,\ 1979]{sand79} Sandage, A.,
Tammann, G.A., and Yahil, A. 1979, \apj , 
232, 352.

\bibitem [Shectman et al.,\ 1996]{she96} Shectman, S.A.,
Landy, S.D., Oemler, A., Tucker, D.L., Lin, H.,
Kirshner, R.P., Schechter, P.L. 1996, \apj ,  470, 172
\bibitem [Schechter 1976]{sch76} Schechter, P. 1976, \apj , 203, 297.
\bibitem [Small et al., \ 1997]{sma97} Small, T., Sargent, W.L.W. and Hamilton, D. 1997, \apjs ,
111.1.
\bibitem [Thorstensen et al.,\  1989]{tho89} Thorstensen, J.R.,
 Wegner, G.A., Hamwey, R., Boley, F.,
Geller, M.J., Huchra, J.P., 
Kurtz, M.J. and  Mc Mahan, R.K. 1989,  \aj ,  98, 1143. 



\bibitem [Thortensen et al.,\ 1995]{tho95} Thorstensen, J.R., 
Kurtz, M.J., Geller, M.J., Ringwald, F.A.,
and Wegner, G. 1995, \aj , 109, 2368.
\bibitem[Tonry and Davis 1979]{ton79} Tonry, J. and Davis, M. 1979, \aj ,
43, 393.
\bibitem [Vettolani et al.,\  1997]{vet97} Vettolani, G., Zucca, E.,
Zamorani, G., Cappi, A., Merighi, R., Mignoli, M., Stirpe, G.M., 
MacGillivray, H., Collins, C., Balkowski, C., Cayette, V., Maurogordato, S.,
Proust, D., Chincarini, G., Guzzo, L., Maccagni, D., Scaramella, R.,
Blanchard, A., and Ramella, M.,  
1997, \aap, in press.

\bibitem [Willmer et al.,\ 1996]{wil96} Willmer, C.N.A.,
Koo, D.C., Ellman, N., Kurtz, M.J., Szalay and A.S.
1996, \apjs ,  104, 199.
\bibitem [Yahil et al.,\  1991]{yah91} Yahil, A.,
Strauss, M.A., Davis, M., and Huchra, J.P. 1991,
\apj , 372, 380.
\bibitem [Zucca et al.,\ 1997]{zuc97} Zucca, E.,
Zamorani, G., Vettolani, P., Cappi, A., Merighi, M.,
Stirpe, G.M., MacGillivray, H., Collins, C., Balkowski, C., Cayette, V.,
Maurogordato, S., Proust, D., Chincarini, G., Guzzi, L., Maccagni, D.,
Scaramella, R., Blanchard, A., and Ramella, M. 1997, \aap , submitted. 

\end{thebibliography}
\end{document}